\documentclass[conference, 10pt]{IEEEtran}
\usepackage{cite}
\usepackage{amsmath,amssymb,amsfonts}
\usepackage{algorithmic}
\usepackage{graphicx}
\usepackage{textcomp}
\usepackage{multirow}
\usepackage[table,xcdraw]{xcolor}
\usepackage{booktabs}
\usepackage[hidelinks]{hyperref}
\usepackage[caption=false,font=footnotesize]{subfig}
\usepackage{colortbl}
\usepackage{multirow}
\usepackage{hhline}
\usepackage{color}
\usepackage{tabularray}
\usepackage{array}
\usepackage{makecell}

\begin{document}

\title{Site-Specific Outdoor Propagation Assessment and Ray-Tracing Analysis for Wireless Digital Twins}

\author{
 \IEEEauthorblockN{Morteza Ghaderi Aram\IEEEauthorrefmark{1}, Hao Guo\IEEEauthorrefmark{1}\IEEEauthorrefmark{2}, Mingsheng Yin\IEEEauthorrefmark{2}, and Tommy Svensson\IEEEauthorrefmark{1}
\IEEEauthorblockA{\IEEEauthorrefmark{1}Department of Electrical Engineering, Chalmers University of Technology, Gothenburg, Sweden\\\IEEEauthorrefmark{2}Tandon School of Engineering, New York University, Brooklyn, NY, USA\\
}
\{aramg, hao.guo, tommy.svensson\}@chalmers.se, my1778@nyu.edu
}}

\maketitle

 \begin{abstract}
Digital twinning is becoming increasingly vital in the design and real-time control of future wireless networks by providing precise cost-effective simulations, predictive insights, and real-time data integration. This paper explores the application of digital twinning in optimizing wireless communication systems within urban environments, where building arrangements can critically impact network performances. We develop a digital twin platform to simulate and analyze how factors such as building positioning, base station placement, and antenna design influence wireless propagation. The ray-tracing software package of Matlab is compared with Remcom Wireless InSite. Using a realistic radiation pattern of a base transceiver station (BTS) antenna, ray tracing simulations for signal propagation and interactions in urban landscapes are then extensively examined. By analyzing radio heat maps alongside antenna patterns, we gain valuable insights into optimizing wireless deployment strategies. This study highlights the potential of digital twinning as a critical tool for urban planners and network engineers.

 \end{abstract}


\section{Introduction}
Usually, network simulators only provide stochastic or simplified hybrid methods for the characterization and simulation of electromagnetic (EM) propagation \cite{boban2014geometry}. Ray-tracing (RT) methods, on the other hand, can provide higher precision under certain conditions, at the cost of demanding computational power and a strong dependency on the quality of the 3D CAD model of the environment \cite{yun2015ray}. Formulated based on a high-frequency approximation (optical ray) to the Maxwell equations, RT methods describe the EM field as a set of propagating, reflecting, diffracting, and scattering rays over environmental elements. In fact, modern ray tracers are all built upon an efficient combination of 3 well-established EM modeling methods \cite{weinmann2009utd}:

\begin{itemize}
    \item Shooting-and-Bouncing-Rays (SBR) algorithm on the basis of geometrical optics (GO) and Fermat's least traveling time principle.
    \item Source-based calculation of scattered field strength using physical optics (PO).
    \item Diffraction calculation on the basis of the uniform theory of diffraction (UTD).
\end{itemize}

Despite the steady introduction of new methods and the extensive body of literature that uses RT techniques, there is still an evident lack of available open-source tools and libraries. Most RT research software is not released or is only commercially licensed. Among currently available open-source simulators is Opal \cite{egea2021opal} which is a ray-launching-based deterministic RF propagation simulator, implemented in C++ with NVIDIA Optix \cite{parker2010optix}, a library and application framework for high-performance ray-tracing on GPU. An application-specific simulator, RadioPropa\footnote{\url{https://github.com/nu-radio/RadioPropa}}, and a general-purpose wireless simulator, PyLayers\footnote{\url{https://pylayers.github.io/pylayers/}}, can also provide some functionality similar to Opal but lack other features such as environment generation tools. When it comes to commercially available software packages, some proprietary tools such as Wireless InSite Remcom \cite{WIRem} and Matlab's Ray-Tracer available via its communication toolbox \cite{Mathworks} offer integrated tools with 3D environment generation and propagation simulators. However, the procurement of CAD inputs in these software packages is not always straightforward and trivial, especially for scenarios with a high level of detail. In this study, we enhance the CAD preparation process for RT-based simulations in a digital twin and compare some of the functionalities available in these two software packages. Based on the conclusions drawn from this comparison, we then extensively explore the potential of Wireless InSite for coverage analyses in a specific urban environment.


\section{Materials \& Methods} 
Both MathWorks and Remcom offer specialized tools for modeling EM signal propagation in rural, urban, and indoor environments. Matlab’s ray tracing model finds line-of-sight (LOS) and non-line-of-sight (NLOS) paths using either the SBR method \cite{ling1989shooting} or the image method. The \texttt{propagationModel} function can be used to specify the appropriate method based on a trade-off between the desired accuracy and the simulation runtime.

Wireless InSite from Remcom also includes several propagation models, both empirical and deterministic such as X3D, Urban Canyon, COST-Hata, and Walfisch-Ikegami. We chose to use the X3D high-fidelity ray-tracing model which provides full control over geometry shapes and/or transmitter/receiver height. According to its reference manual, the X3D solver is supposed to provide accurate predictions from approximately $100$ MHz to about $30$ GHz, thus extending the validity of its propagation calculations up to millimeter wave frequencies. The model considers the effects of reflections, diffraction, transmissions, and atmospheric absorption, and can output such channel characteristics as path loss, delay spread, direction of arrival/departure, and the complex impulse response. 

\subsection{Urban CAD Model, Vegetation, and Terrain}
\begin{figure*}
    \centering
    \subfloat[]{\includegraphics[width=0.635\columnwidth]{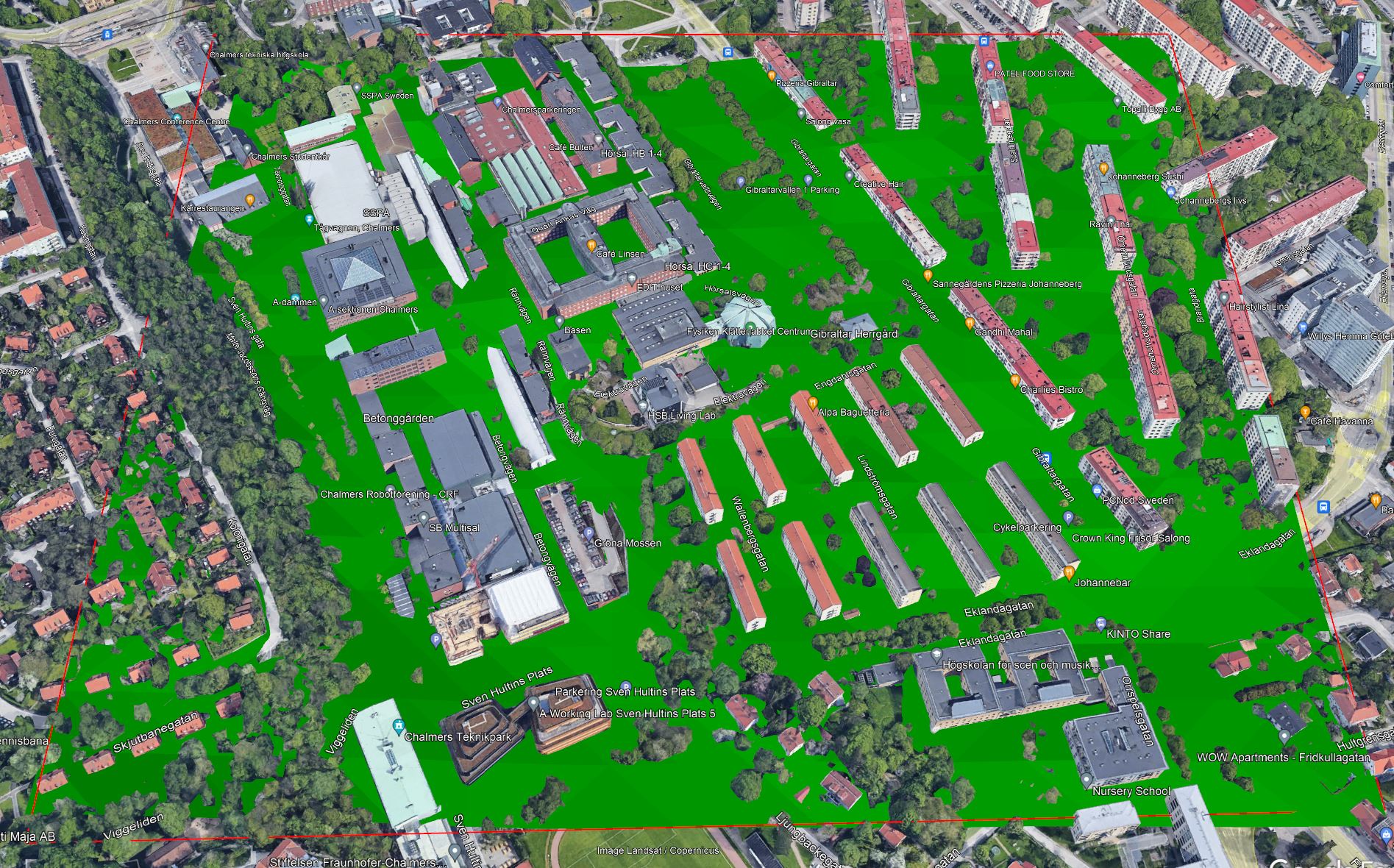}%
    \label{Fig_GeoData_GE}}
    \hfil
    \subfloat[]{\includegraphics[width=0.635\columnwidth]{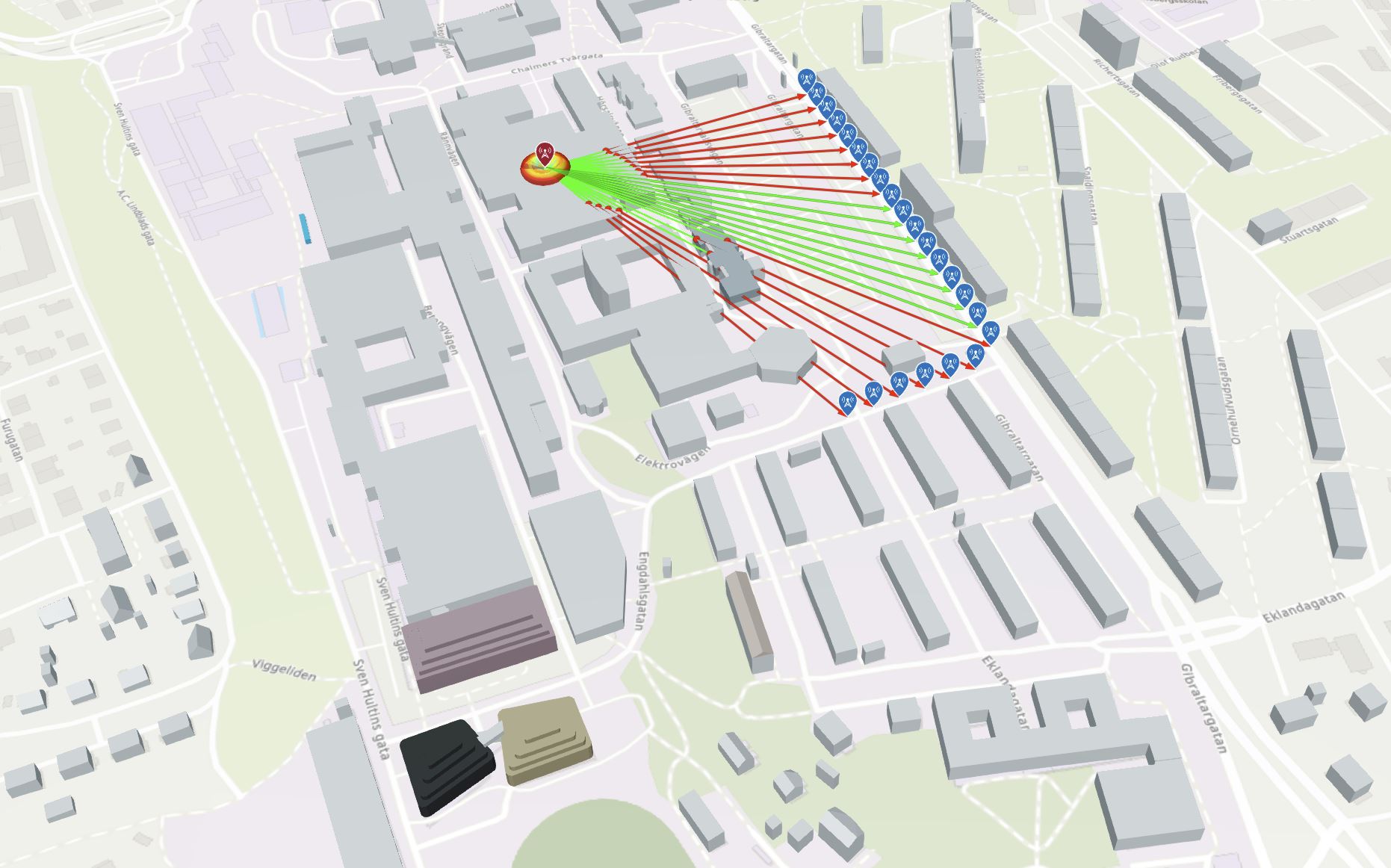}%
    \label{Fig_GeoData_Mat}}
    \subfloat[]{\includegraphics[width=0.7\columnwidth]{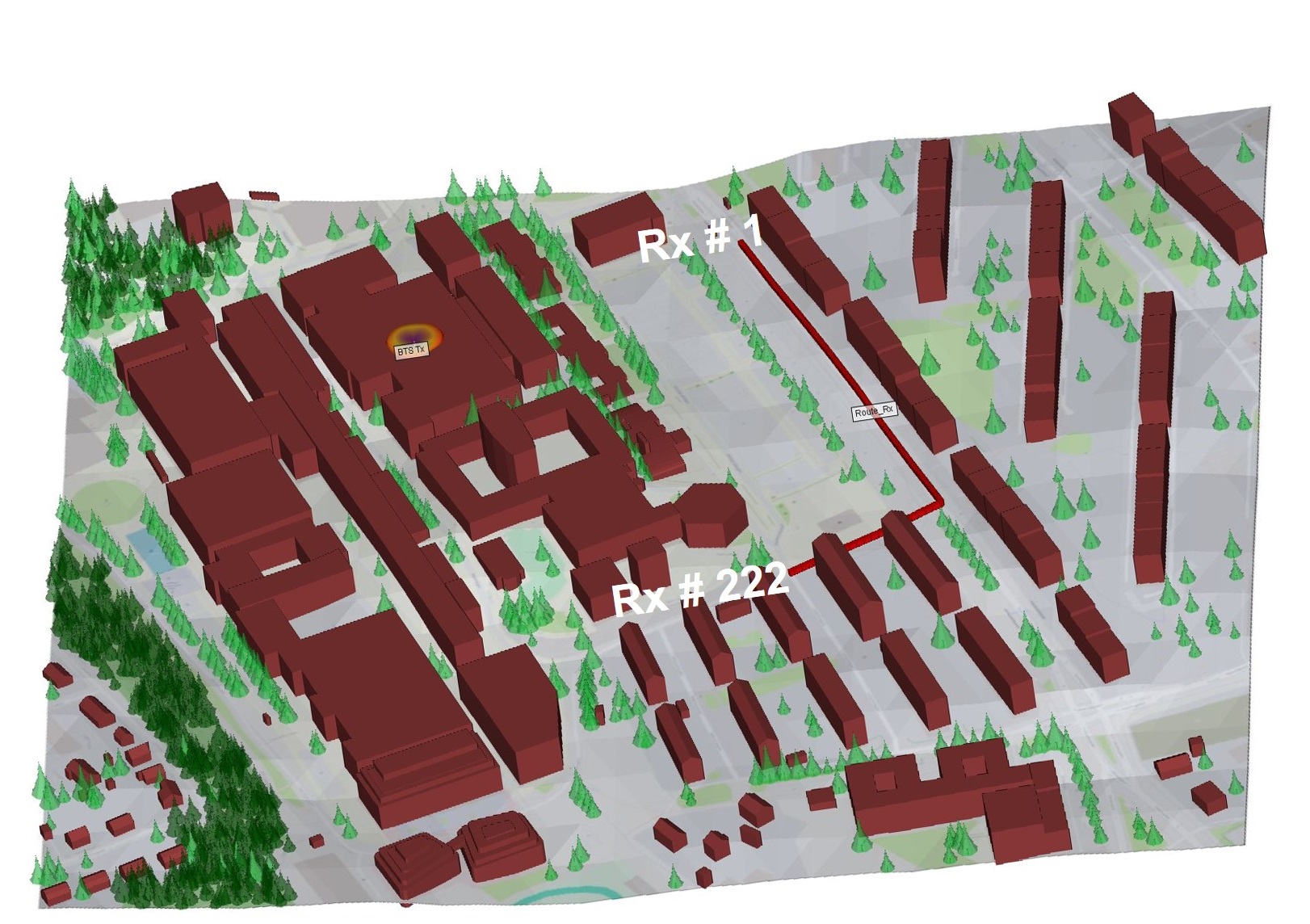}%
    \label{Fig_GeoData_WI}}
    \caption{(a) Google Earth Pro's zoom-in on the highlighted area which is meant to be imported into the ray-tracers. (b) Extracted area from OpenStreetMap by MATLAB which also shows the LoS (green) and NLoS (red) links between the transmitter and the receivers along a route. (c) The extracted model using Blosm which is imported into Wireless InSite; the 3D CAD model of the buildings and trees as well as the Tx antenna's radiation pattern and Rx antennas' route are also shown.}
    \label{Fig_GeoData}
\end{figure*}

As stated above, RT requires the capability to build 3D models of the scenario as a pre-processing step, but current simulators do not usually provide it. In fact, despite the availability of open real-world map data such as OpenStreetMap \cite{OpenStreetMap}, tools to convert Geodata to 3D scenes suitable for propagation simulations are scarce. As a step towards addressing the issue, the goal here is to have a platform that allows for a quick and intuitive generation of customized 3D environments, usually extracted from OpenStreetMap (OSM) data for a specific city district. The area under study here is the premises around the Electrical Engineering department of Chalmers University of Technology, henceforth called the `E2 parking lot' in this study, in downtown Gothenburg, Sweden, with roughly $0.5\times{}0.5$ square-kilometer as shown and highlighted in Fig. \ref{Fig_GeoData_GE}.

For Matlab-based RT, the CAD preparation process is automatic and starts by reading in an OSM file using the \texttt{siteviewer} function. All the metadata and the geospatial information associated with the scenery is then at the user's fingertips by calling the \texttt{readgeotable} function. Matlab uses this information to assign material properties to the objects in the scene based on ITU-R P.2040 for building materials and ITU-R P.527 for terrain materials. Figure \ref{Fig_GeoData_Mat} shows the scenario of the E2 parking lot extracted and used by Matlab's ray tracer. 

When it comes to wireless InSite simulations, there is a very handy add-on to the open-source software package Blender, called Blosm\footnote{\url{https://github.com/vvoovv/blosm}}, which streamlines the CAD preparation process to a large extent. It can download and import real-world terrain data with an approximate resolution of $30$ m, build the 3D CAD model of the buildings from the OpenStreetMap dataset, and finally place the CAD model on top of the imported terrain. A wide assortment of available options for building heights, number of floors, and roof shapes (e.g. flat or gabled) are available in the software to create the final scene. Plus, the premium version of this toolbox can also import forests and single trees as 3D objects. Shown in Fig. \ref{Fig_GeoData_WI} is the 3D CAD model of the E2 parking lot extracted using this procedure and subsequently used for Wireless InSite Simulations with material properties summarized in Table \ref{tab:Mat}. 

\definecolor{DoveGray}{rgb}{0.4,0.4,0.4}
\definecolor{MineShaft}{rgb}{0.129,0.129,0.129}
\definecolor{Silver}{rgb}{0.749,0.749,0.749}
\begin{table}
\caption{Relative permittivity and conductivity of the materials used in 
 Remcom simulations.\begin{tiny}
     \\(V) and (H) stand for vertical polarization and horizontal polarization, respectively.
 \end{tiny}}\label{tab:Mat}
\centering
\resizebox{0.75\columnwidth}{!}{
\begin{tblr}{
  cells = {fg=MineShaft},
  row{2} = {Silver},
  row{4} = {Silver},
  row{7} = {Silver},
  row{8} = {Silver},
  cell{5}{1} = {r=2}{},
  cell{5}{2} = {r=2}{},
  cell{5}{3} = {r=2}{},
  cell{5}{4} = {r=2}{},
  cell{7}{1} = {r=2}{},
  cell{7}{2} = {r=2}{},
  cell{7}{3} = {r=2}{},
  cell{7}{4} = {r=2}{},
  vlines = {DoveGray},
  vline{2} = {1-5,7}{black},
  hline{1-5,7,9} = {-}{},
  hline{6,8} = {5}{},
}
\textbf{Materials}                                           & \bf{Re}$\boldsymbol{\{\epsilon_r\}}$ & $\boldsymbol{\sigma}$ \textbf{(S/m)} & \textbf{Color}    & \textbf{Attenuation}  \\
\textbf{Brick}                                               & 
  4.44
            & 
  0.001
           & 
  Red
           & 
  --
                \\
\textbf{Concrete}                                            & 
  7
               & 
  0.015
           & 
  White
         & 
  --
                \\
{\textbf{ITU Medium Dry\textit{}}\\\textbf{Earth @ 3.5 GHz}} & 
  13.23
           & 
  0.27
            & 
  Gray
          & 
  --
                \\
{\textbf{Dense\textit{}}\\\textbf{Foliage}}                  & 
  --
              & 
  --
              & 
  Light Green
   & 
  (V) 1 dB/m
        \\
                                                             &                     &                     &                   & 
  (H)
  1 dB/m
      \\
{\textbf{Dense Deciduous}\\\textbf{Forest in leaf}}          & 
  --
              & 
  --
              & 
  Dark Green
    & 
  (V) 1.11 dB/m
     \\
                                                             &                     &                     &                   & 
  (H)
  1.64 dB/m
   
\end{tblr}}
\end{table}

\begin{figure}
\centering
\includegraphics[width=\columnwidth]{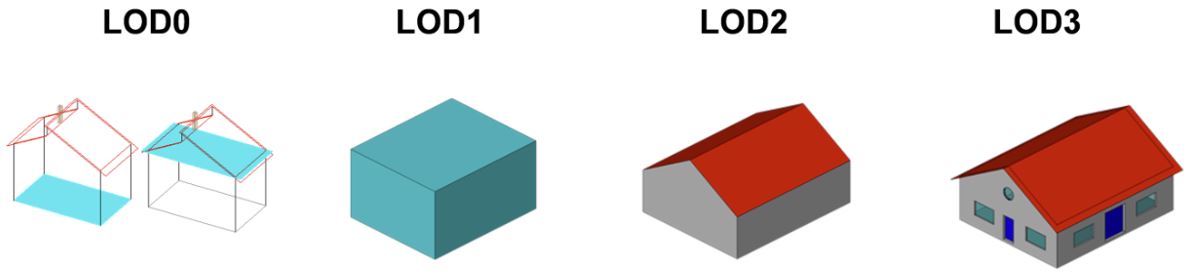}
\caption{Different LODs and their complexity in the CityGML standard.}
\label{Fig_LoD}%
\end{figure}

Regarding the complexity of the imported 3D CAD model for buildings and rooftop shapes, we refer to the CityGML standard, a conceptual model and exchange format for the representation and storage of virtual 3D city models. It facilitates the integration of urban Geodata for smart cities and digital twins for a variety of applications, ranging from, e.g., landscape planning to mobile telecommunications. According to this model, there exist four levels of detail (LOD) for outdoor city models that progressively grow in complexity from the buildings' footprint (LOD$0$) to their block model (LOD$1$), the coarse exterior (LOD$2$), and finally, the fine exterior (LOD$3$), as can be seen in Fig. \ref{Fig_LoD}. For our current study, LOD$2$ seems to provide sufficient details. LOD$3$ will however need to be implemented in a future study if better control on building facades for reconfigurable intelligent surface (RIS) integration and/or deployment is to be practiced. 

\subsection{BTS Antenna Far-Field Radiation Pattern}
For an accurate coverage prediction, radio network planning tools need to in general consider a number of crucial base station parameters such as height, orientation, antenna downtilt, and antenna radiation pattern, among which antenna patterns are of special importance. Indeed, from the 1\textsuperscript{st}-generation (1G) to the 5\textsuperscript{th}-generation (5G) mobile networks, base transceiver station (BTS) antennas have been the edge element in the air interface towards the mobile terminal in all communication systems. They are mainly designed and deployed in cellular communication systems in two basic forms: omnidirectional and pencil-beam (directional) antennas \cite{farasat2021review}. For low-capacity and extended coverage scenarios such as rural areas, the omnidirectional option is preferred while the pencil-beam directional pattern better serves a targeted coverage area in demand of a higher capacity. 

From 2G BTSs onwards, sectorization has been a common practice in which the omnidirectional cell is often divided into three sectors of $120^{\circ}$ each. In typical urban deployments, base stations are usually located in elevated positions such as rooftop or telecommunication towers. Ideally, the radio signal propagates above rooftops, gets diffracted at rooftop edges, and is guided by the street canyons (through multiple reflections on buildings and corner diffractions) until it reaches the user terminal \cite{zhao2017channel, rodriguez2014base}. 

\begin{figure}[t!]
    \centering
    \subfloat[]{\includegraphics[width=0.5\columnwidth]{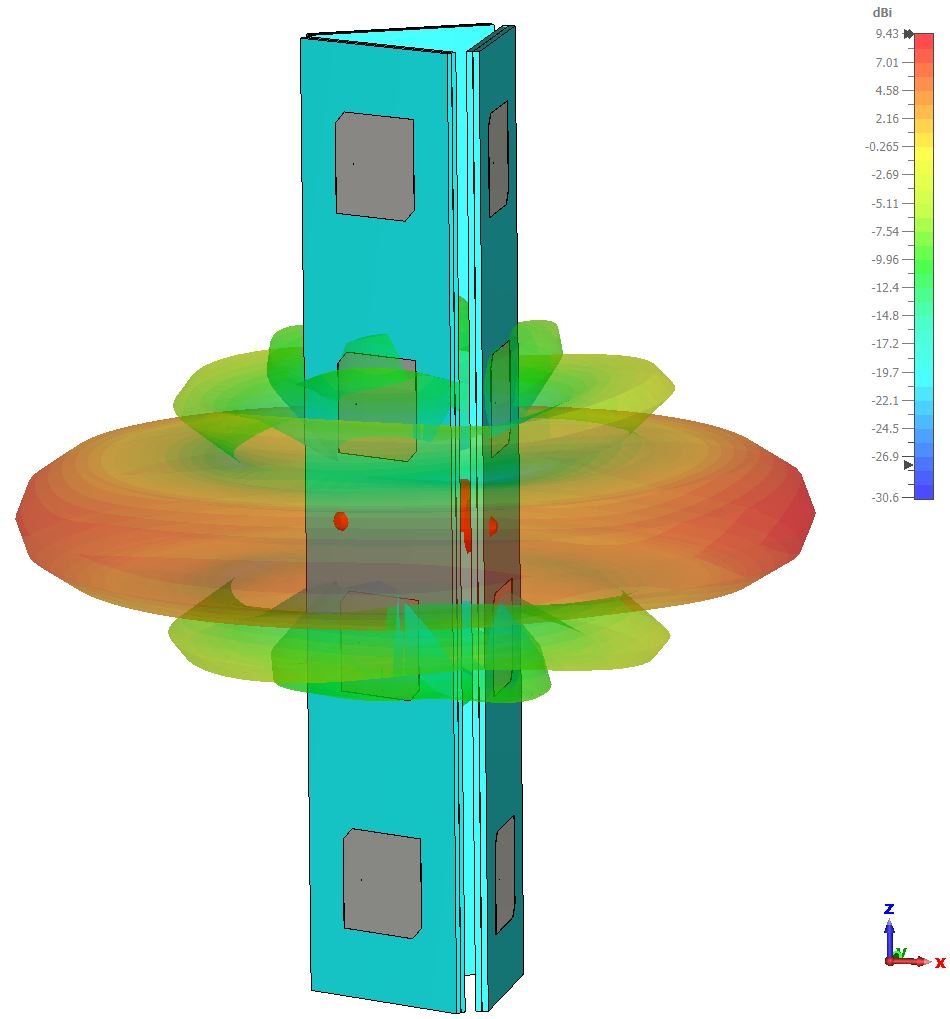}%
    \label{Fig_FFpat}}
    \hfil
    \subfloat[]{\includegraphics[width=0.5\columnwidth]{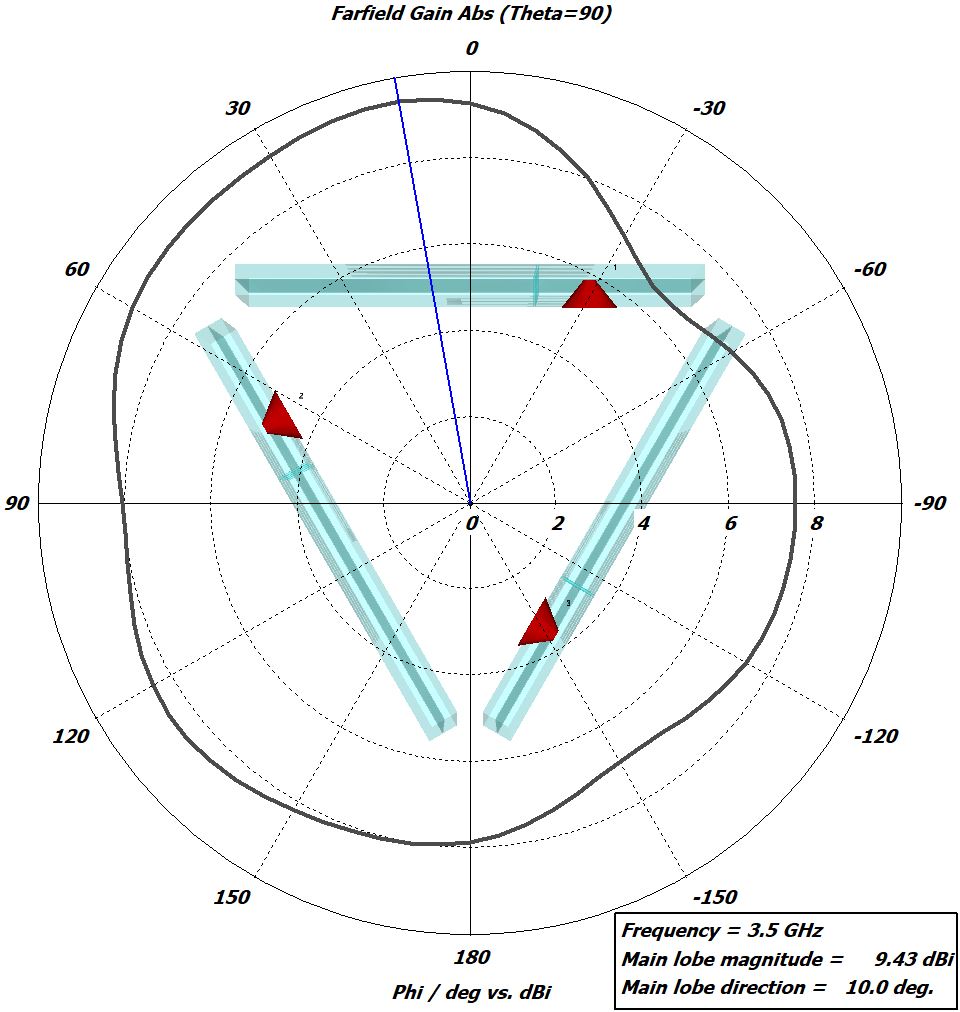}%
    \label{Fig_2DFFpat}}
    \caption{CST Simulation setup and the Far-field pattern for the designed tri-sectoral BTS antenna (a) 3D pattern (b) 2D cut in the azimuth plane for $\theta=90^{\circ}$.}
    \label{Fig_BTS_Ant}
\end{figure}

The designed sub-$6$ GHz antenna at the NR n$78$ frequency band with the center frequency of $3.5$ GHz is shown in Fig. \ref{Fig_FFpat} which is an example of multi-sectoral panel BTS antennas. 
It combines three panels of $4\times{}1$ patch arrays in a tri-sectoral configuration to achieve an omnidirectional pattern as shown in Fig. \ref{Fig_2DFFpat}. The far-field radiation pattern of the design is calculated using the finite integration technique (FIT) of CST Microwave Studio Suite\textsuperscript{\textregistered}. The pattern is then imported into Matlab and Wireless InSite as a user-defined antenna element for subsequent ray-tracing simulations. 

\subsection{Solvers Parameters and Setup}
The parameters and the simulation setups for both solvers are summarized in Table \ref{tabSpec}. In its current implementation in Matlab, the SBR method considers only reflections and edge diffractions and supports up to ten path reflections and one edge diffraction for coverage analysis. In the following simulations, we set the number of diffractions in Matlab to its maximum, i.e. $1$, while setting $4$ as the maximum number of reflections. Higher values for this parameter, up to $6$ as is the case in Wireless InSite, were also tried, but they made the simulations prohibitively long by placing stringent memory demands on the personal workstation handling the simulations. Plus, we observed that signal strengths after $4$ rounds of reflection were too weak to have a significant impact on the predicted coverage plots. 

Capable of running on GPUs and using multi-threading in multi-core processors, X3D solver of Wireless InSite is a full 3D ray-tracing model that uses exact path correction (EPC) to overcome some of the challenges faced by the traditional SBR method in its path-finding algorithm \cite{WIRem}. 
The X3D solver can handle a maximum number of reflections, transmissions, and diffractions up to 30, 8, and 3, respectively. Users can also choose between the Weissberger model or the generic attenuation model in terms of dB/m when it comes to foliage modeling. 

\begin{table}[t]
\caption{Simulation Setup for Ray-Tracing in both solvers}\label{tab:SolvParam}
\begin{center}
\resizebox{0.9\columnwidth}{!}{
\begin{tabular}{ccc}
\toprule
 & \multicolumn{2}{c}{\textbf{Values}} \\
\textbf{Parameters} & Wireless InSite & Matlab \\
\midrule
Max \# of Reflection & $6$ & $4$ \\
Max \# of Diffraction & $2$ & $1$ \\
Max \# of Transmission & $0$ & $0$ \\
Frequency & $3.5$ GHz &  $3.5$ GHz\\
Tx Antenna & \makecell{Omni-directional \\ (User-defined Tri-sectoral)} & \makecell{Omni-directional \\ (User-defined Tri-sectoral)} \\
Rx Antenna & \makecell{Half-wavelength \\Vertical Dipole} & Isotropic \\
\makecell{Tx Height from \\ Rooftop/Ground} & $1.5$ (m)/ $18.5$ (m) &  $1.5$ (m)/ $18.5$ (m)\\
\makecell{Rx Height from \\ Ground} & $1.6$ (m) &  $1.6$ (m)\\
Tx Power & $30$ dBm $=1$ W & $30$ dBm $=1$ W\\
Foliage Model & Weissberger &  NA\\
\bottomrule
\end{tabular}}
\label{tabSpec}
\end{center}
\end{table}

\section{Results \& Discussion}

\subsection{Solvers Parameters Effects on the Predicted RSS}
\label{RoutSec}
\begin{figure}[htp]
\centering
\includegraphics[width=\columnwidth]{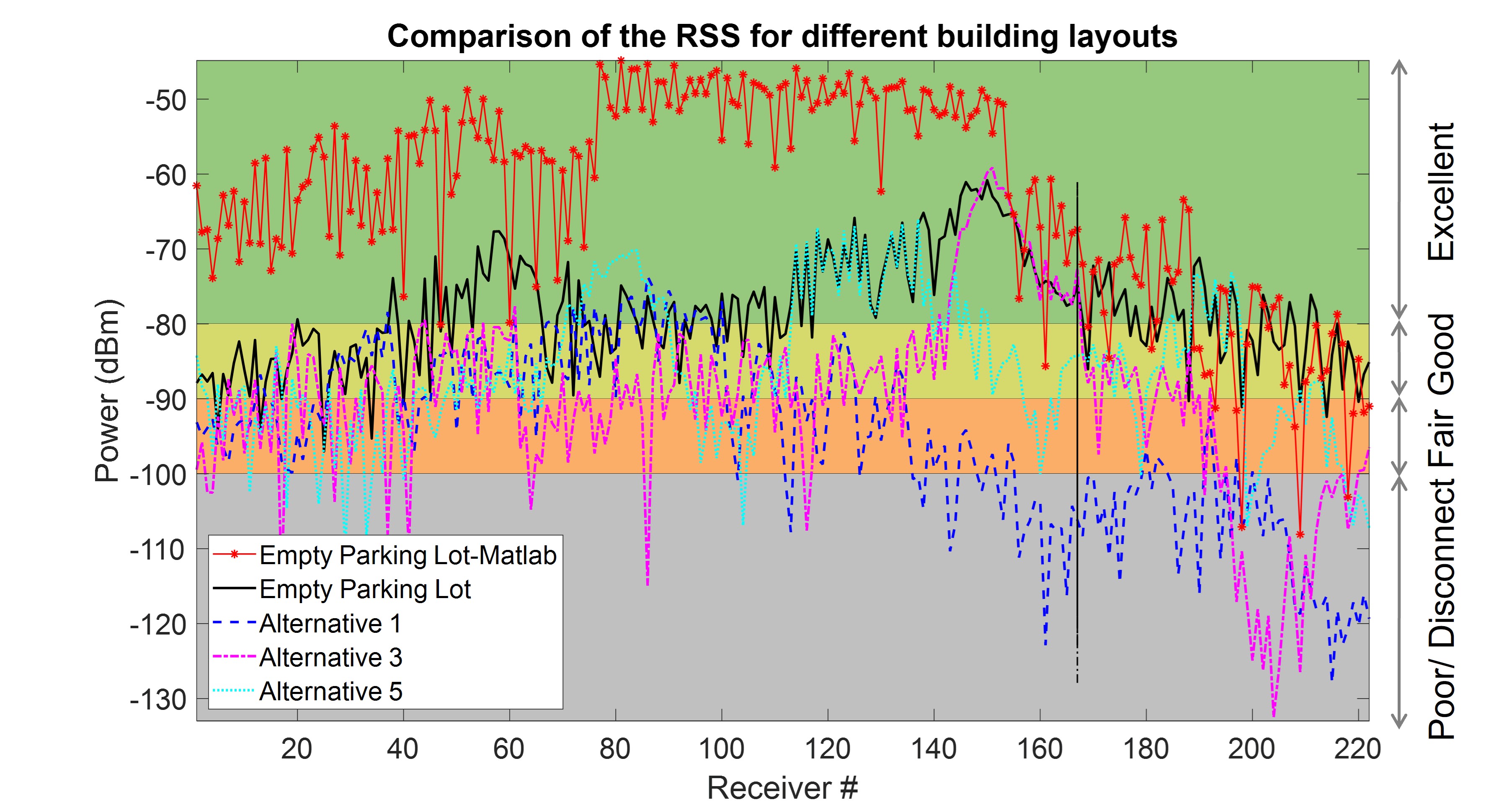}
\caption{RSS along the chosen route predicted by both Matlab and Wireless InSite for the empty parking lot. The predictions for the added three building options shown with colored dashed lines were only made using Wireless InSite.}
\label{Fig_Pr_RxRoute}%
\end{figure}

As depicted in Figs. \ref{Fig_GeoData_Mat} and \ref{Fig_GeoData_WI}, a linear array of $222$ receivers along an L-shaped route was chosen for the first part of this study. For the sake of comparison between the two solvers, the predicted received signal strength (RSS) along the route in the empty parking lot scenario was reported in Fig. \ref{Fig_Pr_RxRoute} as the solid red and black lines for Matlab and Wireless InSite, respectively. The Matlab prediction is on average $15$-$20$ dB higher than that of Wireless InSite in the first part of the path until the route starts to turn around the corner at the receiver number $167$. The predictions of the two solvers are much closer in the second part of the path, i.e., from the corner receiver up to the end receiver. We speculate that the disagreement partly stems from the fact that there is a line of trees situated in front of the first part of the route which is taken into account by Wireless InSite while being ignored by Matlab. The lower maximum diffraction limit in Matlab might be another contributing factor to this noticeable discrepancy between the two sets of results. 

Overall, we encountered the following two drawbacks and/or limitations in the current implementation of the Matlab raytracer. Firstly, it does not extract any information about trees and forests, and hence the propagation loss due to the interaction of EM waves with foliage cannot be captured in its simulations. Secondly, to the best of our knowledge, there is no straightforward way of incorporating STL files into geographic data once the coordinate system is set in Matlab. This of course hampers further modifications or the addition of new buildings/objects to the scene in Matlab. 

\begin{figure}[htp]
    \centering
    \subfloat[]{\includegraphics[width=0.5\columnwidth]{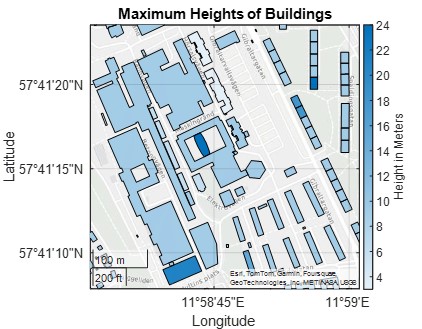}%
    \label{Fig_Heights}}
    \subfloat[]{\includegraphics[width=0.35\columnwidth]{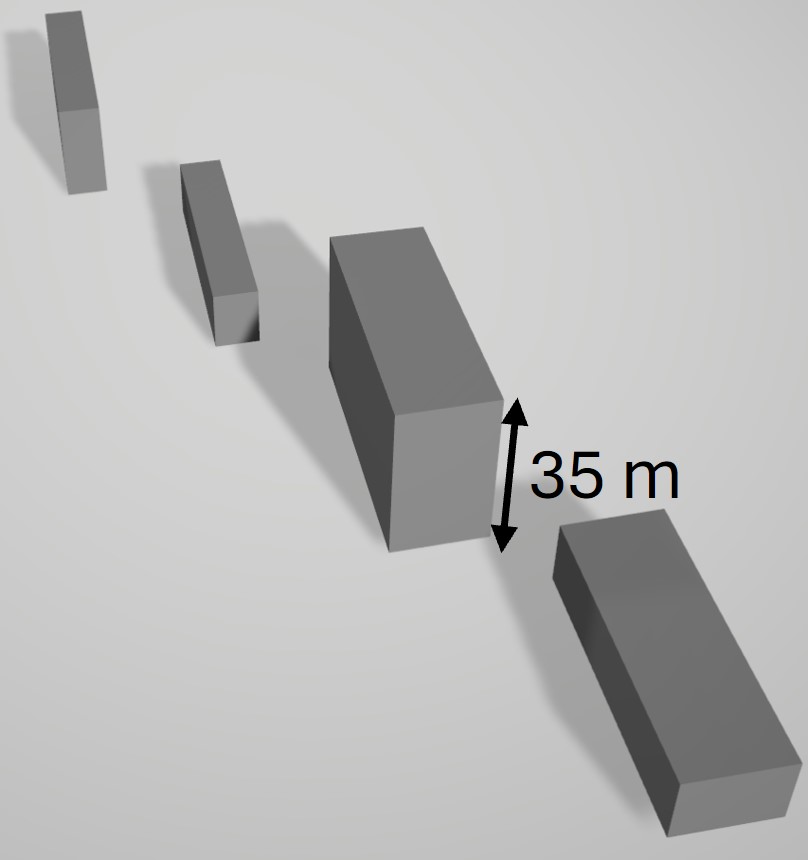}%
    \label{Fig_Alt1H}}
    \\
    \hspace{0.9cm}
    \subfloat[]{\includegraphics[width=0.3475\columnwidth]{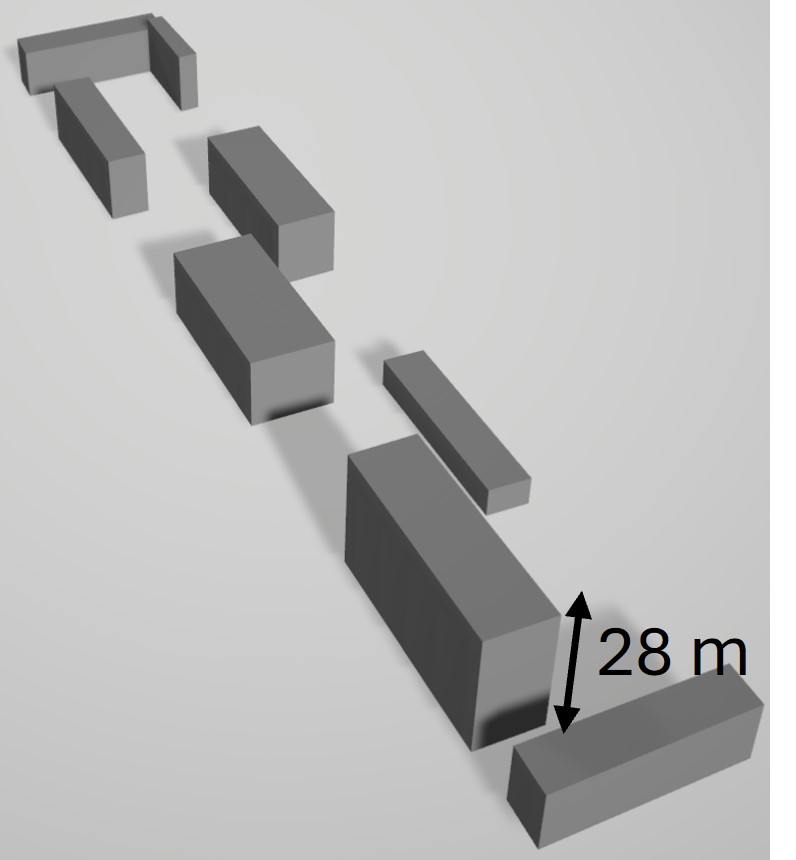}%
    \label{Fig_Alt3H}}
    \hspace{0.2cm}
    \subfloat[]{\includegraphics[width=0.3575\columnwidth]{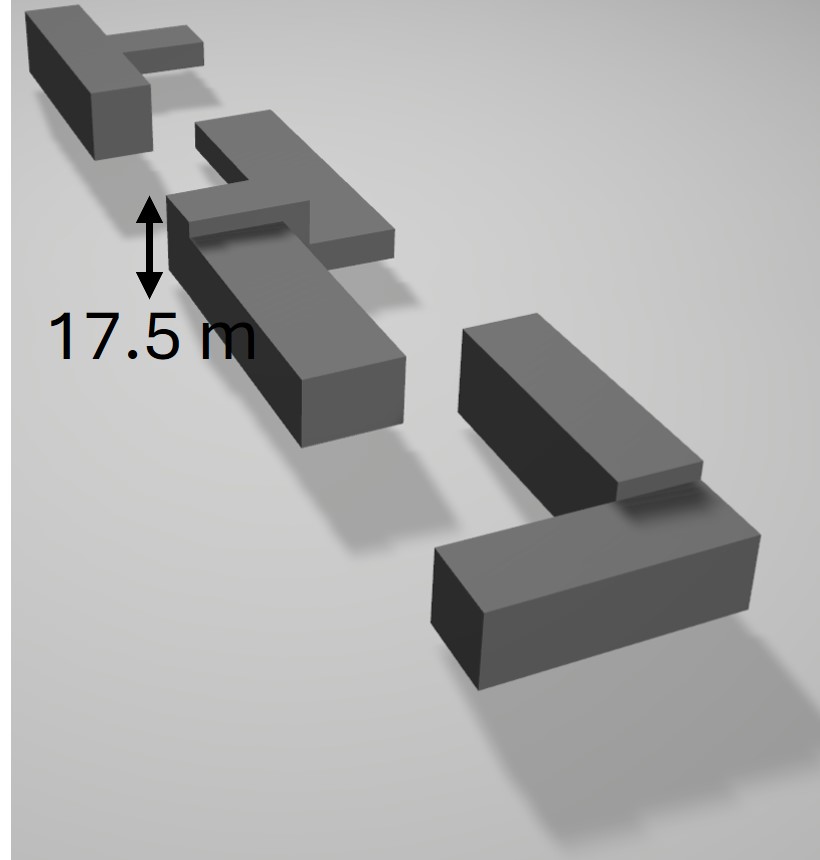}%
    \label{Fig_Alt5H}}
    \caption{Maximum heights of the buildings (a) in the surrounding area of the empty parking lot in front of Chalmers' E2 department and (b) alternative $1$, (c) alternative $3$, and (d) alternative $5$ of the three building layouts suggested by a recent study \cite{gonzalez2024towards} to fit into the parking space.}
    \label{Fig_Alterns}
\end{figure}

For instance, we study the impact of urban building designs on the RSS by fitting three concrete building layouts suggested in a recent study \cite{gonzalez2024towards} into the parking lot as exemplified in Fig. \ref{Fig_Alterns}. Due to the above-mentioned limitations in Matlab, the RSS results for the three building options were only computed by Wireless InSite and reported in Fig. \ref{Fig_Pr_RxRoute} as the colored dash lines. Their means and standard deviations are also summarized in Table \ref{tab:Statistics}. In light of this comparison and given the fact that Wireless InSite provides a more flexible solver setting with less demanding computational resources, in the next section we chose to continue the coverage analysis on the three building layouts only with Wireless InSite.

\definecolor{Silver}{rgb}{0.749,0.749,0.749}
\definecolor{Black}{rgb}{0, 0, 0}
\definecolor{MineShaft}{rgb}{0.129,0.129,0.129}
\definecolor{Concrete}{rgb}{0.949,0.949,0.949}
\begin{center}
\begin{table}
\caption{Mean \& Standard Deviation (SD) of the signal strength along the Chosen Route}\label{tab:Statistics}
\centering
\resizebox{\columnwidth}{!}{
\begin{tblr}{
  cells = {fg=MineShaft},
  row{2} = {Concrete},
  hlines,
  vlines = {Silver},
  vline{2} = {-}{Black},
}
\textbf{Scenarios}  & \textbf{Empty Lot} & \textbf{Alternative 1} & \textbf{Alternative 3} & \textbf{Alternative 5} \\
\textbf{Mean (dBm)} & 
  -78.34
         & 
  -94.8
              & 
  -90.12
             & 
  -86.38
             \\
\textbf{SD (dB)}    & 
  7.27
           & 
  11.68
              & 
  12.62
              & 
  9.55
               
\end{tblr}}
\end{table}
\end{center}

\subsection{Coverage Analysis}
\label{CovSec}
\begin{figure*}[htp]
\centering
\includegraphics[width=1.25\columnwidth]{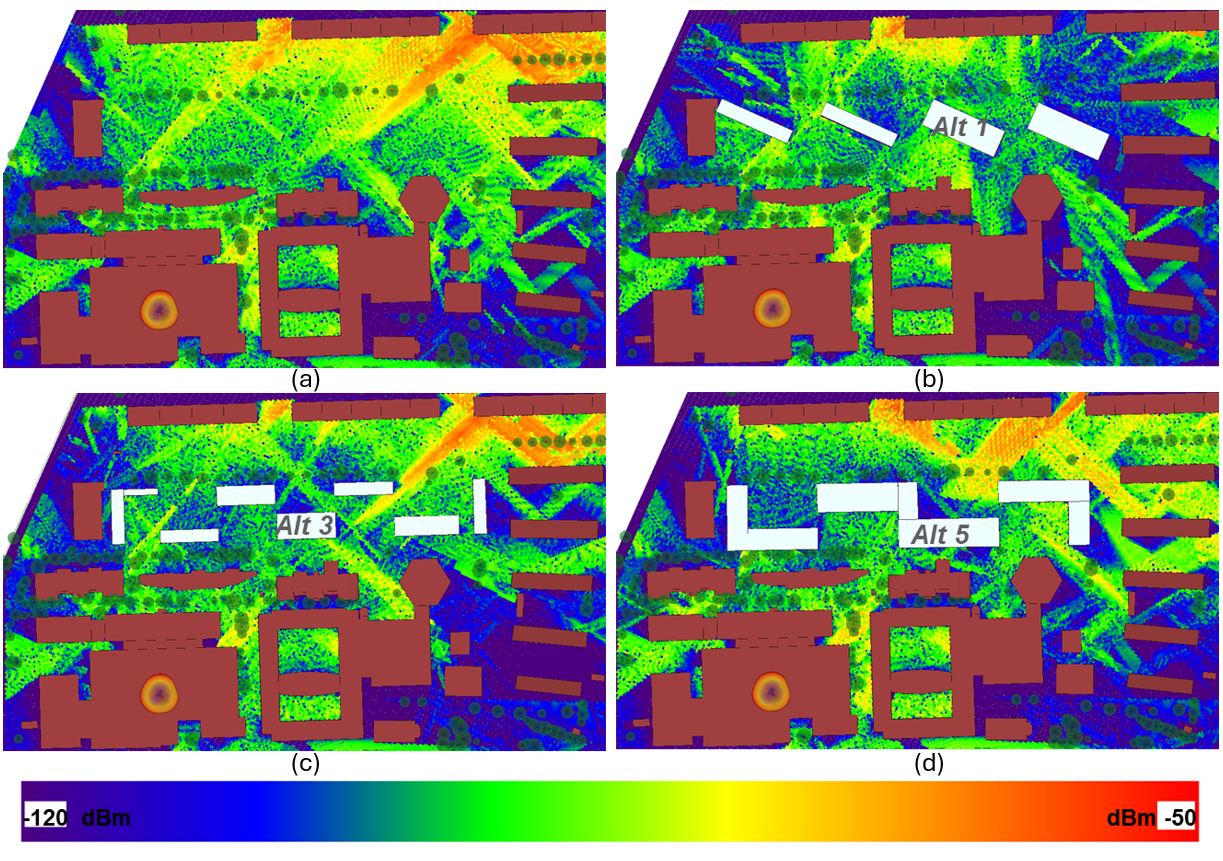}
\caption{Predicted coverage maps at $3.5$ GHz for (a) the empty parking lot and (b) alternative $1$, (c) alternative $3$, and (d) alternative $5$ for the three options of new building layouts made of concrete and shown in Fig. \ref{Fig_Alterns}.}
\label{Fig_Pr_X3d_Altern}%
\end{figure*}

\begin{figure}[htp]
\centering
\includegraphics[width=0.75\columnwidth]{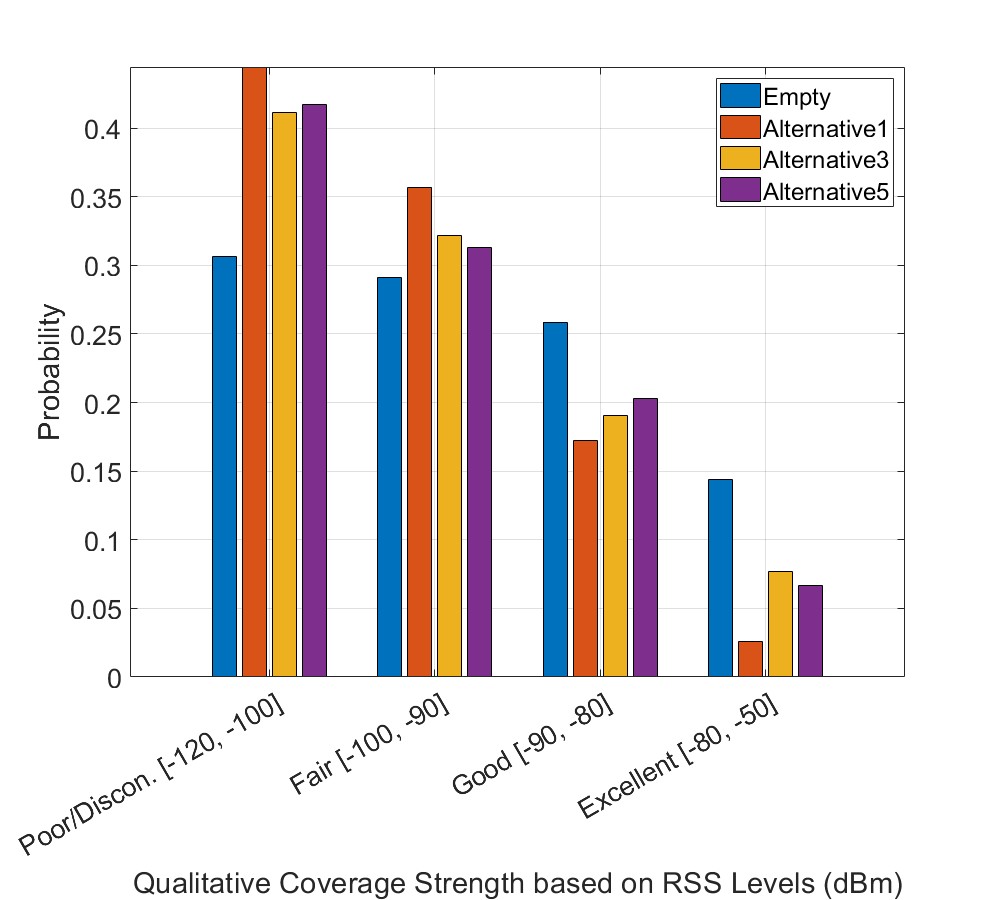}
\caption{Histograms of the four scenarios under study extracted from their respective coverage maps.}
\label{Fig_RSSHisto}%
\end{figure}

Figure \ref{Fig_Pr_X3d_Altern} shows the heat maps predicted by Wireless InSite for the empty E2 parking lot scenario as well as the three building layout scenarios. Through a visual inspection, one can see that the building arrangments and their relative heights with respect to the transmit antenna can have a huge impact on the power distribution pattern in the cell.
In an attempt to quantify the difference among these coverage patterns, we proceeded to report in Fig. \ref{Fig_RSSHisto} the histogram of power distributions for the four scenarios under investigation. The four descriptive categories labeled as excellent, good, fair, and poor are the same as the ones annotated and color-coded on top of Fig. \ref{Fig_Pr_RxRoute}. The histogram clearly shows that the empty parking lot has a higher percentage of excellent and good coverage followed by alternatives $5,~3,$ and finally $1$. In this way, we can conclude that the first building layout most severely degrades the overall coverage quality in the cell.


\section{Conclusion}
In the first part of this study, two ray-tracing software packages were investigated and compared in order to assess their potential integration in a wireless digital twin platform under development. Specific comments and remarks on the current limitations of the solvers were made and it was shown that the foliage and higher-order diffraction effects were important and should not be neglected. Since Wireless InSite provides the capability of foliage modeling and higher order diffraction limit, it was the package of choice for the coverage analysis in the second part of this study, where three different building layouts were investigated. The predicted coverage maps and the extracted histograms stress the fact that urban planning and wireless network deployment are highly intertwined and should be considered conjointly, especially in the digital twin platforms for 5G/6G and beyond networks.  

\section*{Acknowledgements}
This work has been funded in part, by Chalmers AoA Transport project ”DT6GV", and by the Swedish Research Council (VR grant 2023-00272). The authors would also like to extend their gratitude to Remcom for providing a license for the Wireless InSite package.

\bibliographystyle{IEEEtran}
\bibliography{reference}

\end{document}